\documentclass[prl,article,twocolumn,preprintnumbers,superscriptaddress,amsmath,amssymb,aps,longbibliography]{revtex4-1}

\usepackage{graphicx}% Include figure files
\usepackage{dcolumn}% Align table columns on decimal point
\usepackage{bm}% bold math
\usepackage[dvipsnames,table]{xcolor}
\usepackage{hyperref}

\hypersetup{colorlinks,
			citecolor      = niceblue, 
			linkcolor       = niceblue, 
			urlcolor        = niceblue, 
			anchorcolor = niceblue}
\definecolor{niceblue}{rgb}{0.1,0.2,0.6}

\newcommand{\oGW}{\Omega_\text{GW}}

\newcommand{\Trh}{T_\text{rh}}
\newcommand{\mdm}{m_{\text{DM}}}
\newcommand{\Tmax}{T_\text{max}}
\newcommand{\rgw}{\rho_\text{GW}}

\begin{document}

%%%%%%%%%%%%%%%%%%%%%%%%%%%%%%%%%%%%%%%%%%%%%%%%%%%%%%%%%%
\title{Probing Gravitational Dark Matter with Ultra-high Frequency Gravitational Waves}
%%%%%%%%%%%%%%%%%%%%%%%%%%%%%%%%%%%%%%%%%%%%%%%%%%%%%%%%%%
\author{Yong Xu}
\email{yonxu@uni-mainz.de}
\affiliation{{\it PRISMA$^+$} Cluster of Excellence and Mainz Institute for Theoretical Physics\\
Johannes Gutenberg University, 55099 Mainz, Germany}
%%%%%%%%%%%%%%%%%%%%%%%

% \date{\today}% It is always \today, today,

\begin{abstract}
The evidence for the existence of dark matter (DM) is compelling, yet its nature remains elusive. A  minimal scenario involves DM interacting solely through gravity. However, the detection would be extremely challenging.  In the early Universe, such DM can be unavoidably generated via annihilation of particles in the standard model (SM) thermal plasma.  It is known that the SM thermal plasma also produces gravitational waves (GWs). In this study, we establish a simple connection between the amplitude of thermal GWs and the properties of pure gravitational DM.  Notably, future GW experiments in the ultra-high frequency regime have the potential to shed light on the mass and spin of pure gravitational DM.
\end{abstract}

\maketitle

%%%%%%%%%%%%%%%%%%%%%%%%%%%%%%%%%%%%%
\section{Introduction}
%%%%%%%%%%%%%%%%%%%%%%%%%%%%%%%%%%%%%

Observations indicate that non-baryonic dark matter (DM) constitutes approximately 85\% of the total matter content of the Universe~\cite{Bertone:2016nfn, Cirelli:2024ssz, Balazs:2024uyj}. Despite compelling evidence for the existence of DM--such as galactic rotation curves, the Bullet Cluster, and cosmic microwave background (CMB) observations--the nature of DM remains unknown. The possible mass range of DM spans an extraordinary range, from $10^{-21}~\text{eV}$ to $10^{40}~\text{g}$, covering more than 90 orders of magnitude~\cite{Cirelli:2024ssz}. Depending on the mass scale of DM as well as its couplings with visible sectors, the production mechanisms vary, necessitating distinct detection strategies.

In this work, we consider a minimal scenario in which DM interacts solely via gravity with Standard Model (SM) particles. Precise measurements of light element abundances indicate that thermal equilibrium must have been established in the early Universe for successful Big Bang Nucleosynthesis (BBN) \cite{Cyburt:2015mya}. In this context,  DM can be produced in the early Universe through annihilations of SM thermal plasma particles mediated by massless gravitons~\cite{Garny:2015sjg, Tang:2016vch, Tang:2017hvq, Garny:2017kha, Bernal:2018qlk}.  Due to its pure gravitational interactions with visible matter, experimentally detecting such DM is highly challenging. Nevertheless, gravitational waves (GWs) offer a promising avenue to probe this scenario, as we will demonstrate in this work. It is also  known that the SM thermal plasma emits gravitons, thereby sourcing GWs
\cite{Ghiglieri:2015nfa, Ghiglieri:2020mhm, Ringwald:2020ist, Klose:2022knn, Klose:2022rxh, Ringwald:2022xif, Ghiglieri:2022rfp, Muia:2023wru, Drewes:2023oxg, Ghiglieri:2024ghm,  Bernal:2024jim}. Since the production of DM and GWs share the same source, the latter is related to the former. This is in the same spirit as relating the DM abundance with the baryon asymmetry of the Universe (BAU)~\cite{Boucenna:2013wba}. In the case under consideration, the GW spectrum is related to the DM abundance, where the later depends on the mass and spin of DM particles. These correlations enable probing the pure gravitational DM scenario using GWs. In the context of GW and DM, we note that  GWs  might act as a source for fermionic DM production, as discussed recently in Refs.~\cite{Maleknejad:2024ybn, Maleknejad:2024hoz}.

The goal of this article is to demonstrate the minimal and inevitable relationship between the GW spectrum and pure gravitational DM produced from the SM thermal bath. Here is the outline of the rest of the article. In the next section, we offer the setup. Then, we briefly revisit the production of gravitational DM and GWs from the SM thermal plasma. Following that, we present the main result—the connection between properties of pure gravitational DM and the GW spectrum. Finally, we conclude with a summary of the article.

\section{The Setup}\label{sec:setup}
We assume a minimal setup with the following action 
\begin{align}\label{eq:action}
S \supset \int d^4 x\sqrt{-g} \left[ \frac{M_P^2}{2}R+ \mathcal{L}_\text{SM} + \mathcal{L}_\text{DM}  \right]\,,
\end{align}
where $g$ corresponds to the determinant of the metric $g_{\mu \nu}$, and $R$ the Ricci scalar, $M_P\equiv 1/\sqrt{8\pi G_N}$ is  the reduced Planck mass. The standard model Lagrangian is denoted as $\mathcal{L}_\text{SM} $, and $\mathcal{L}_\text{DM}$ includes the mass and kinetic terms for DM. We assume that there is no other interaction except for gravitation for DM. Expanding the metric $g_{\mu \nu}$ around the Minkowski metric $\eta_{\mu \nu}=(+,-,-,-)$, we have \cite{Donoghue:1994dn}
\begin{align}\label{eq:expansion}
g_{\mu \nu} =\eta_{\mu \nu} + \frac{2}{M_P}\, h_{\mu \nu} \,,
\end{align}
where $h_{\mu \nu}$ denotes the massless spin-2 graviton field.  Using Eq.~\eqref{eq:expansion}, one has $\sqrt{-g} \simeq 1+h/M_P$ with $h\equiv  h^{\mu}_{\mu} $ corresponding to the trace of the graviton field, which is zero in the transverse and traceless gauge.  With $g_{\mu \nu}$ in Eq.~\eqref{eq:expansion}, it follows that the contravariant form:
\begin{align} \label{eq:expansion2}
g^{\mu \nu}\simeq \eta^{\mu \nu} - \frac{2}{M_P}\, h^{\mu \nu} + \cdots \,,
\end{align} 
where $\cdots $ encodes higher orders of $\frac{1}{M_P}$. Using these expansions in Eq.~\eqref{eq:action}, we obtain the following effective couplings \cite{Donoghue:1994dn}:
\begin{align}\label{eq:effective}
\sqrt{-g} \mathcal{L} \supset \frac{1}{M_P} h_{\mu \nu} \sum_i T_i^{\mu \nu}\,,
\end{align}
where $T_i^{\mu \nu}$ corresponds to the energy-momentum tensor for a particle species $i$, including SM and DM. Eq.~\eqref{eq:effective} implies that DM and GW can be produced from SM particle annihilations.

%%%%%%%%%%%%%%%%%%%%%%%%%%%%%%%%%%%%%
\section{Cogenesis of  Dark Matter  and Gravitational Wave}\label{sec:DM}
%%%%%%%%%%%%%%%%%%%%%%%%%%%%%%%%%%%%%
\begin{figure}[!ht]
\def\sepf{0.22}
\centering
\includegraphics[scale=\sepf]{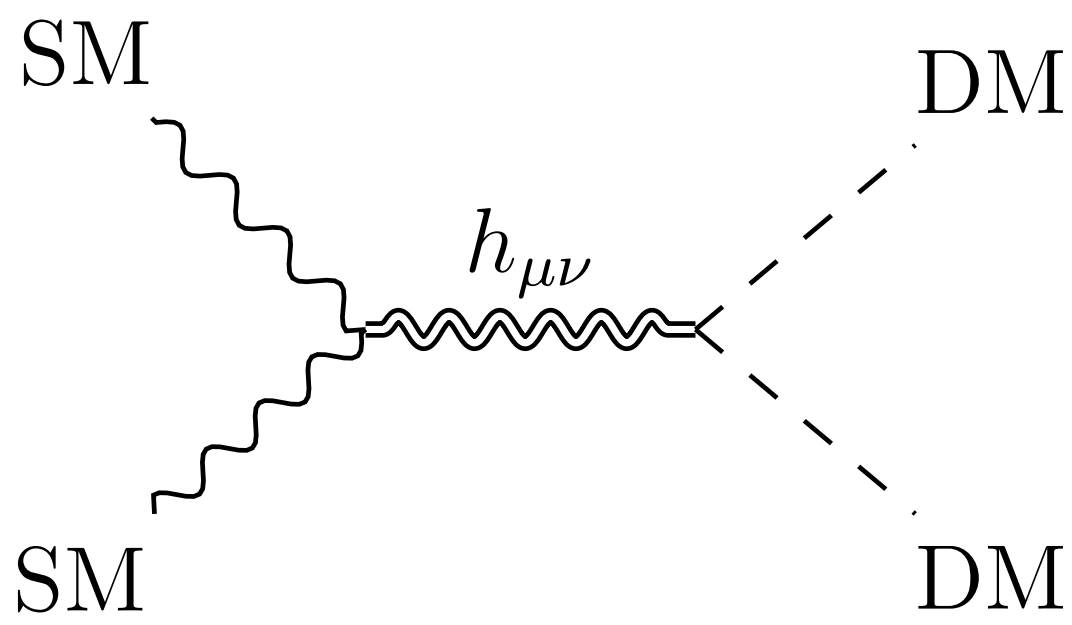}
\includegraphics[scale=\sepf]{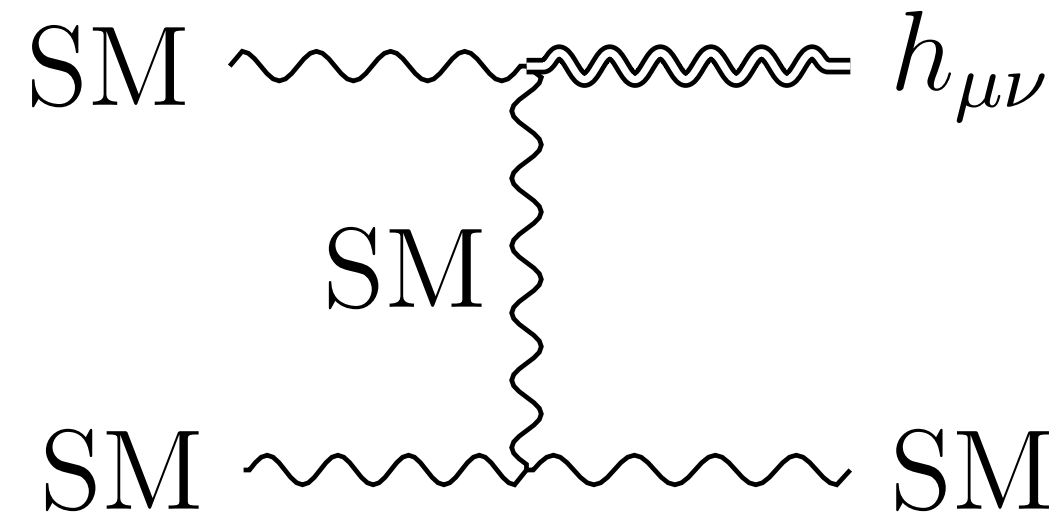}\\
\caption {Feynman diagrams for  cogenesis of DM  (left) and GW (right) from SM thermal bath.}
\label{fig:diagram}
\end{figure} 
%%%%%%%%%%%%%%%%%%%%%%%%%%%%%%%%%%%%%
As discussed above in Eq.~\eqref{eq:effective}, DM can be generated through the annihilations of SM particles in the thermal plasma as shown in the left panel of Fig.~\ref{fig:diagram}. The interaction rate density takes a form  $\gamma_{\text{DM}} \simeq \alpha \frac{T^8}{M_P^4}$, where the value of $\alpha$ depends on the spin of the DM particles \cite{Bernal:2018qlk}.  Similarly to the production of DM, massless spin-2 gravitons can also be generated via annihilation of SM particles,  as shown by the example in the right panel of Fig.~\ref{fig:diagram}. Here, we have focused on the leading processes with a single graviton production. Double gravitons production is possible but suppressed by an extra $1/M^2_P$ in the cross section \cite{Ghiglieri:2022rfp, Ghiglieri:2024ghm}. The production rate density of gravitons takes the form $\gamma_{h} \propto  \frac{T^6}{M_P^2}$ ~\cite{Ghiglieri:2015nfa, Ghiglieri:2020mhm, Ringwald:2020ist, Bernal:2024jim}.  It is important to note that the temperature under consideration is well below $M_P$, such that gravitons do not thermalize. After production, gravitons would propagate throughout the Universe, forming  a stochastic GW background.

The evolution of DM and graviton number densities is governed by the Boltzmann equation:
\begin{align}
\dot{n}_i+ 3H n_i=\gamma_i\,,
\end{align}
where $\dot{n}$ denotes $dn/dt$ with $t$ being cosmic time, $H$ Hubble rate, and $i$ accounts for DM and graviton. By solving the Boltzmann equations, we can obtain the relic abundance of DM and GW at present. The DM abundance is given by
%%%%%%%%%%%%%%%%%%%%%%%%%%%%%%%%%%%%%%%%%%%%%%
\begin{align}\label{OmegaDM}
\Omega_{\text{DM}} h^2 \simeq 0.12 \left(\frac{\alpha}{3 \cdot 10^{-3}}\right)  \left(\frac{\Trh}{10^{14}~\text{GeV}}\right)^3  \left(\frac{\mdm}{10^9~\text{GeV}}\right)\,,
\end{align}
%%%%%%%%%%%%%%%%%%%%%%%%%%%%%%%%%%%%%%%%%%%%%%%
where  $\alpha =1.9\times 10^{-4}\,, 1.1 \times 10^{-3}\,, ~\text{and}~2.3\times 10^{-3}$ for DM with spin $s=0$, $s=1/2$, and $s=1$, respectively \cite{Bernal:2018qlk}. Note that the differences among these values arise from the distinct kinematics of DM particles in the final states. Eq.~\eqref{OmegaDM} applies for DM with mass $\mdm \lesssim \Trh$, where the reheating temperature $\Trh$ denotes the temperature at the beginning of radiation domination. 

Similarly, one can also obtain the amount of GW at present, which is usually quantified as $\oGW \equiv \frac{1}{\rho_c}\frac{d \log \rgw}{d \log f}$. Here, $\rho_c \equiv 3\,H_0^2 M_P^2$ denotes the critical energy densities  with $H_0$ denoting the Hubble parameter  at present, $\rgw$ and $f$ correspond to the GW energy density and frequency at present, respectively. The GW spectrum can be written as 
\begin{align}\label{OGW1}
    \Omega_\text{GW}h^2\simeq 8.6 \cdot 10^{-11} \left(\frac{\Trh}{10^{14}~\text{GeV}}\right) \left(\frac{f}{10^{11}~\text{Hz}}\right)^3 \hat{\eta}(f)\,,
\end{align}
where $\hat{\eta}$ encodes  the sources for the GW production ~\cite{Ghiglieri:2015nfa, Ghiglieri:2020mhm, Ringwald:2020ist, Bernal:2024jim}. Note that $\hat{\eta}$ includes a Boltzmann suppression term for gravitons with energies significantly exceeding the temperature of the thermal bath. After production, gravitons redshift in the same way as temperature, leading to a peak frequency of the GW spectrum similar to that of CMB photons, with $ f \sim \mathcal{O}(100)~\text{GHz} $. 

Since both DM and GW production arise from the thermal plasma, they are interconnected through the reheating temperature $\Trh$, sharing a common origin. Using Eq.~\eqref{OmegaDM}, one can rewrite Eq.~\eqref{OGW1} as
\begin{widetext}
\begin{align}\label{OGW2}
    \Omega_\text{GW}h^2&\simeq 8.6 \cdot 10^{-11} \left(\frac{\Omega_{\text{DM}} h^2 }{0.12} \right)^{1/3} \left(\frac{\alpha}{3\cdot 10^{-3}}\right)^{-1/3}   \left(\frac{\mdm}{10^9~\text{GeV}}\right)^{-1/3} \left(\frac{f}{10^{11}~\text{Hz}}\right)^3 \hat{\eta}(f)\,.
\end{align}
\end{widetext}
Eq.~\eqref{OGW2} is one of the main results of this article, providing the mathematical formulation of the consequence for the unavoidable cogenesis of DM and GWs as illustrated in Fig.~\ref{fig:diagram}. It demonstrates how the GW spectrum is related with the mass and spin properties of pure gravitational DM. We remind the reader again that the value of $\alpha$ depends on the DM spin.

Several comments are in order before closing this section. We begin by noting that the above analysis assumes a thermal history in which the Universe remains radiation-dominated for $T \leq \Trh$, a minimal scenario supported by precise measurements of light element abundances~\cite{Cyburt:2015mya}. If an early matter-dominated phase or a prolonged reheating epoch precedes radiation domination, the maximum temperature of the thermal bath, $\Tmax$, can be (significantly) higher than the reheating temperature $\Trh$~\cite{Giudice:2000ex}.  In such cases, both DM and GW production could be enhanced due to the ultraviolet freeze-in behavior during reheating, i.e., in the regime $\Trh \leq T \leq \Tmax$. Consequently, Eq.~\eqref{OGW2} acquires corrections that depend on $\Tmax$. For typical reheating scenarios, where a heavy, matter-like inflaton oscillates in a quadratic potential—such as those arising from Starobinsky inflation~\cite{Starobinsky:1980te} or polynomial inflation~\cite{Drees:2021wgd, Drees:2022aea}—the maximum temperature can be expressed as~\cite{Xu:2024fjl}   $\Tmax=\Trh \left(3/8\right)^{2 / 5}\left[H_{\text{inf}}/H(\Trh)\right]^{1 / 4}$, where $H(\Trh)$ is the Hubble parameter at $T = \Trh$, and $H_{\text{inf}}$ denotes the Hubble parameter during inflation. Since reheating follows inflation, we have $H_{\text{inf}} > H(\Trh)$, thus explaining why $\Tmax > \Trh$. We have verified that the impact of the reheating phase introduces, however, only a minor correction factor of $\mathcal{O}(1)$ to Eq.~\eqref{OGW2}. Indeed, Eq.~\eqref{OmegaDM} acquires an additional factor of 2~\cite{Bernal:2024ykj}, while Eq.~\eqref{OGW1} receives a logarithmic correction $\log(\Tmax / \Trh) \sim \mathcal{O}(1)$~\cite{Bernal:2024jim}.

Furthermore, if reheating is assumed, both DM and GWs could also be produced through inflaton interactions mediated by gravity. This leads to an alternative scenario in which the cogenesis of DM and GWs is purely sourced by the inflaton. While this extends beyond our current assumption that both DM and GWs originate from a thermal bath, as illustrated in Fig.~\ref{fig:diagram}, we will comment on this possibility. We begin by noting that DM could be generated via pure gravitational inflaton annihilations during reheating~\cite{Chung:1998zb, Chung:1998rq, Ema:2018ucl, Ema:2019yrd, Mambrini:2021zpp, Clery:2021bwz, Kolb:2023ydq}. Similarly, GWs sourced by graviton generation could also arise from pure inflaton annihilation~\cite{Ema:2015dka, Ema:2020ggo, Choi:2024ilx, Xu:2024fjl}. However, the resulting GW spectrum is significantly more suppressed, primarily due to the $1/M^4_P$ suppression in the graviton production cross section. For this reason, testing such a cogenesis scenario using GWs is significantly more challenging compared to the scenario focused on in this work.

%%%%%%%%%%%%%%%%%%%%%%%%%%
\section{Results}\label{sec:GW_DM}
\begin{figure}[!ht]
\def\sepf{0.7}
\centering
\includegraphics[scale=\sepf]{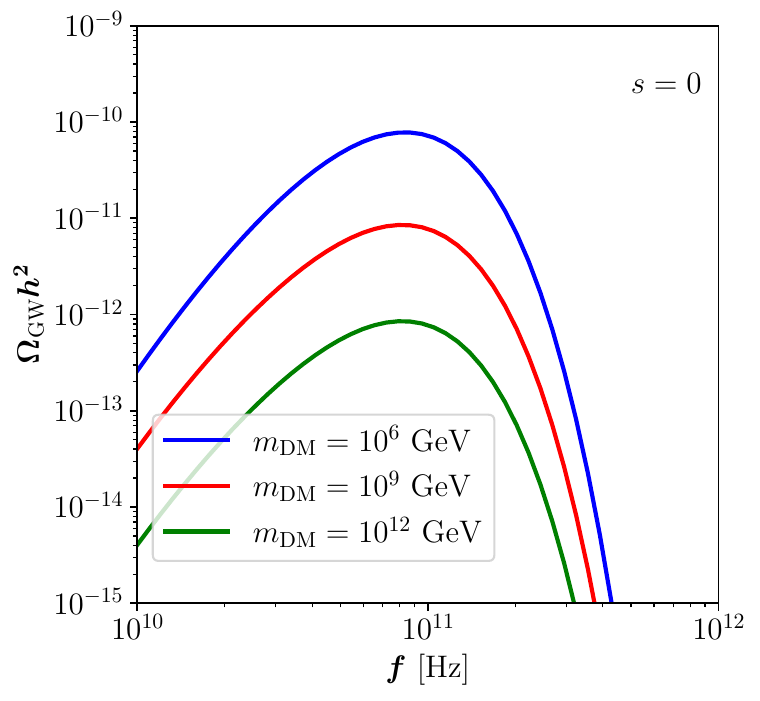}
\includegraphics[scale=\sepf]{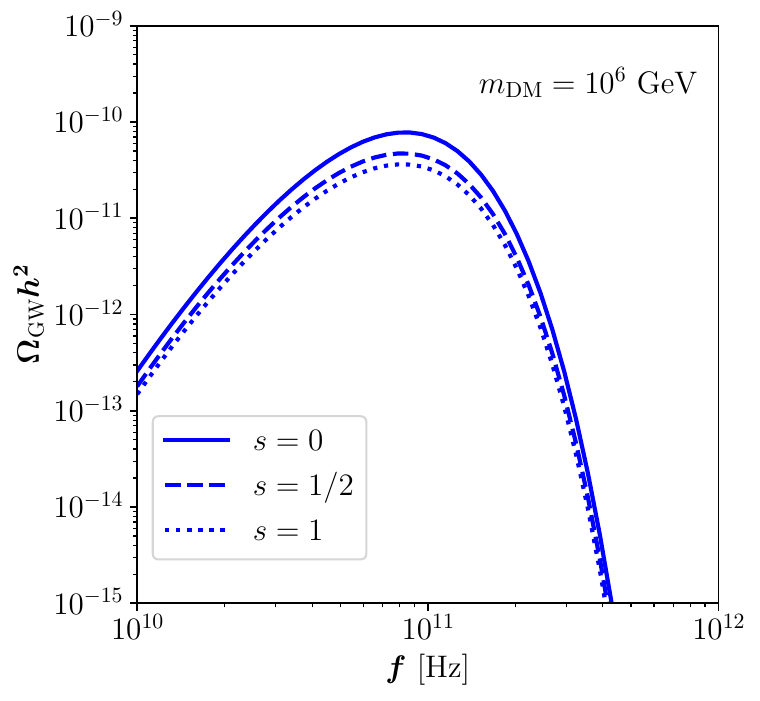}
\caption {GW spectrum as function of frequency $f$, the mass $\mdm$ and spin  $s$ of gravitational DM.}
\label{fig:GW_DM}
\end{figure}

%%%%%%%%%%%%%%%%%%%%%%%%%%
In this section, we present the numerical results, which are illustrated in Fig.~\ref{fig:GW_DM}.

%%%%%%%%%%%%%%%%%%%%%%%%%%%
%%%%%%%%%%%%%%%%%%%%%%%%%%%

In the upper panel of Fig.~\ref{fig:GW_DM}, the GW  spectrum is shown as a function of frequency $f$ and DM mass $\mdm$, with the DM spin fixed at $s=0$, corresponding to $\alpha =1.9 \times 10^{-4}$. The blue, red, and green lines correspond to $\mdm = 10^{6}~\text{GeV}$, $\mdm = 10^{9}~\text{GeV}$, and $\mdm = 10^{12}~\text{GeV}$, respectively.  Due to the redshift effect during reheating, the peak frequency $f_{\text{peak}} $ of the GW spectrum is slightly lower than that in the standard radiation domination case around $ 80~\text{GHz}$ \cite{Bernal:2024jim}.  Imposing $\Omega_{\text{DM}} h^2 = 0.12$ requires a higher reheating temperature $\Trh$ for smaller DM masses (cf. Eq.~\eqref{OmegaDM}), leading to a stronger GW amplitude. This explains why the blue curve lies above the other two. For DM with much smaller mass than the values shown in the upper panel of Fig.~\ref{fig:GW_DM}, a significantly higher $\Trh$ would be needed to reproduce the observed relic abundance. Given the current upper bound on inflationary tensor-to-scalar ratio from cosmic microwave background (CMB) measurements, such as the BICEP/Keck 2018 results~\cite{BICEP:2021xfz}, which constrain $\Trh \lesssim 5.5 \times 10^{15}~\text{GeV}$, we find $\mdm \gtrsim 3.2 \times 10^4~\text{GeV}$ using Eq.~\eqref{OmegaDM}.

In the lower panel of Fig.~\ref{fig:GW_DM}, we fix the DM mass to $\mdm = 10^{6}~\text{GeV}$ and consider different DM spins: $s=0$ (blue solid line), $s=1/2$ (blue dashed line), and $s=1$ (dotted line). The variation in GW spectrum amplitude across different spins suggests that GW measurements with enough resolution may also provide information about the spin of pure gravitational DM particles.

We now connect our results to potential future ultra-high frequency GW experiments. The development of ultra-high frequency GW detection is an active and rapidly expanding area of research, with many new proposals emerging~\cite{Cruise:2012zz, Aggarwal:2020olq, Aggarwal:2025noe}. We demonstrate that measuring the amplitude of GWs in the $f \sim \mathcal{O}(10^{11})~\text{Hz}$ regime can provide novel probe of the mass and spin of pure gravitational DM, assuming it constitutes the entire relic abundance produced from the SM thermal bath. In particular, a null result for $\Omega_{\text{GW}} h^2 \gtrsim \mathcal{O}(10^{-10})$ at $f \sim \mathcal{O}(10^{11})~\text{Hz}$ would exclude the parameter space for pure gravitational DM with $\mdm \lesssim 10^6~\text{GeV}$ (cf. the upper panel of Fig.~\ref{fig:GW_DM}). Conversely, a positive detection might hint at the existence of pure gravitational DM and even could potentially offer information about its spin, provided that GW experiments achieve sufficient resolution (cf. the lower panel of Fig.~\ref{fig:GW_DM}).

%%%%%%%%%%%%%%%%%%%%%%%%%%
\section{Conclusion}\label{sec:conclusion}
%%%%%%%%%%%%%%%%%%%%%%%%%%%
The existence of a thermal bath in the early Universe and dark matter (DM) is well established based on experimental observations. While the evidence for DM's existence is compelling, its nature remains unknown. A pure gravitational DM scenario represents a minimal framework in which DM can be generated via the annihilation of Standard Model (SM) particles in the thermal plasma, which simultaneously emits gravitational waves (GWs).  Due to its purely gravitational interactions, direct detection of such DM is highly challenging. In this work, we establish a novel connection between the GW spectrum and the properties of pure gravitational DM, arising from their inevitable cogenesis from the SM thermal plasma in the early Universe. The main result is summarized by Eq.~\eqref{OGW2} and illustrated in Fig.~\ref{fig:GW_DM}. We demonstrate that future ultra-high-frequency GW experiments could potentially provide a novel probe of both the mass and spin of pure gravitational DM.

%%%%%%%%%%%%%%%%%%%%%%%%%%
\section*{acknowledgments}
%%%%%%%%%%%%%%%%%%%%%%%%%%%
YX thanks Manuel Drees for helpful discussions as well as comments on the draft. The author also wishes to thank Nicolás Bernal and Andreas Mantziris for discussions. YX acknowledges the support from the Cluster of Excellence ``Precision Physics, Fundamental Interactions, and Structure of Matter'' (PRISMA$^+$ EXC 2118/1) funded by the Deutsche Forschungsgemeinschaft (DFG, German Research Foundation) within the German Excellence Strategy (Project No. 390831469).

\bibliographystyle{apsrev4-1}
\bibliography{biblio}

\end{document}